\documentclass[10pt,twocolumn,letterpaper]{article}

\usepackage{style}
\date{}
\usepackage[pagebackref,breaklinks,colorlinks]{hyperref}      

\usepackage{xcolor}
\usepackage{amsmath}
\usepackage{subcaption}
\usepackage{multirow}
\usepackage{arydshln}
\usepackage{booktabs}
\usepackage{adjustbox}

\title{\LARGE \bf
Facial-Expression-Aware Prompting for Empathetic LLM Tutoring
}

\author{\parbox{16cm}{\centering
{\large
Shuangquan Feng$^{1}$ \hspace{0.6cm}
Laura Fleig$^{2}$ \hspace{0.6cm}
Ruisen Tu$^{3}$ \hspace{0.6cm}
Philip Chi$^{3}$ \hspace{0.6cm}
Edmund Bu$^{4,5}$\\[4pt]
Melinda Ozel$^{6}$ \hspace{0.6cm}
Junhua Ma$^{3}$ \hspace{0.6cm}
Teng Fei$^{2}$ \hspace{0.6cm}
Virginia R. de Sa$^{2,7}$
}\\[6pt]
{\normalsize
$^1$ Neurosciences Graduate Program, University of California San Diego, La Jolla, USA\\
$^2$ Department of Cognitive Science, University of California San Diego, La Jolla, USA\\
$^3$ Department of Computer Science and Engineering, University of California San Diego, La Jolla, USA\\
$^4$ Department of Electrical and Computer Engineering, University of California San Diego, La Jolla, USA\\
$^5$ Department of Mathematics, University of California San Diego, La Jolla, USA\\
$^6$ Face the FACS\\
$^7$ Halıcıoğlu Data Science Institute, University of California San Diego, La Jolla, USA
}
}
\thanks{
We are grateful for support from the California Education Learning Lab, seed funding from UC San Diego Social Sciences and the Sanford Institute for Empathy and Compassion, funds from the Hal\i c\i o\u{g}lu Data Science Institute Endowed Chair I, and anonymous donor funds to Cognitive Science as well as hardware funding from NVIDIA, Adobe, and Sony.
}
}

\begin{document}

\maketitle

\begin{abstract}
Large language models (LLMs) enable increasingly capable tutoring-style conversational agents, yet effective tutoring requires sensitivity to learners’ affective and cognitive states beyond text alone. Facial expressions provide immediate and practical cues of confusion, frustration, or engagement, but remain underexplored in LLM-driven tutoring. We investigate whether facial-expression-aware signals can improve empathetic tutoring responses through prompt-level integration, without end-to-end retraining. We build a scalable simulated tutoring environment where a student agent exhibits diverse facial behaviors from a large unlabeled human facial expression video dataset, and compare four tutor variants: a text-only LLM baseline, a multimodal baseline using a random facial frame, and two Action Unit estimation model (AUM)–based methods that either inject textual AU descriptions or select a peak-expression frame for visual grounding. Across 960 multi-turn conversations spanning three tutor backbones (GPT-5.1, Claude Opus 4.5, and Gemini 2.5 Pro), we evaluate targeted pairwise comparisons with five human raters and an exhaustive AI evaluator. AU-based conditioning consistently improves empathetic responsiveness to facial expressions across all tutor backbones, while AUM-guided peak-frame selection outperforms random-frame visual input. Textual AU abstraction and peak-frame visual injection show model-dependent advantages. Control analyses show that this improvement does not come at the expense of worse pedagogical clarity or responsiveness to textual cues. Finally, AI–human agreement is highest on facial-expression-grounded empathy, supporting scalable AI evaluation for this dimension. Overall, our results show that lightweight, structured facial expression representations can meaningfully enhance empathy in LLM-based tutoring systems with minimal overhead. Future work should evaluate these methods using real student facial data collected during authentic learning interactions, and examine whether improved empathetic responsiveness translates into measurable benefits in engagement and learning outcomes.
\end{abstract}

\section{Introduction}
Large language models (LLMs) have enabled increasingly capable conversational systems in domains such as customer support and education \cite{wang2024survey, wang2023aligning, wulf2024utilizing, wang2024large}. However, effective human-aligned interaction often requires sensitivity to users' affective and cognitive states, especially when users provide limited textual feedback \cite{sorin2024large, wang2023emotional}. Multimodal signals can provide complementary cues for such adaptation \cite{baltruvsaitis2018multimodal, karray2008human, duric2002integrating}, and facial expressions are particularly practical indicators of states like confusion, frustration, or engagement \cite{hess1995intensity}. Despite this, facial behavior remains underexplored in LLM-driven interaction frameworks compared to text or audio \cite{hu2022acoustically}.

AI tutoring presents a particularly compelling and intuitive scenario to demonstrate the feasibility and value of incorporating empathy into multimodal LLMs (MLLMs). Effective tutoring depends in part on the tutor’s ability to recognize and respond to learners’ emotional and cognitive states, such as confusion and engagement. Recent work on LLM-based tutoring systems has shown their potential to provide personalized, scalable, and effective educational support \cite{wang2024large,dai2023can,stamper2024enhancing}. In this context, AI tutoring provides a clear and practical domain for examining the feasibility and value of empathetic multimodal interaction. 
In this work, we investigate whether injecting facial-expression-aware signals into tutoring-style LLM responses can improve empathetic responsiveness without end-to-end retraining. We incorporate structured facial representations derived from Action Units (AUs) and compare two lightweight integration strategies: (i) converting AU estimates into concise textual descriptions prepended to a text-only tutor prompt, and (ii) using AU-based saliency to select a peak frame for a multimodal tutor. Through controlled simulated multi-turn tutoring conversations and both human and AI evaluation, we analyze when and how facial-expression-awareness improves empathy in tutoring responses.

\section{Related Work}
\subsection{Affective Intelligence in Education}
AI systems increasingly support education through intelligent tutors, automated grading tools, and adaptive learning platforms. However, many systems primarily focus on cognitive tasks (e.g.
content delivery, problem-solving) 
while overlooking emotional states that significantly influence learning  \cite{woolf2009affect}.
Affective intelligence (the ability of a system to detect, interpret, and respond to human emotions) has thus become increasingly important in the design of educational AI. 
In tutoring systems, recognizing affective cues such as confusion or frustration allows for better-timed and more effective support, enhancing engagement and learning \cite{d2012dynamics}.

\subsection{Facial Expression Recognition}

To effectively detect emotional cues, researchers have explored a variety of multimodal signals. Among them, facial expressions are particularly rich in affective content and often provide the clearest window into emotional states. In educational settings, facial expressions serve as crucial indicators of student engagement and comprehension \cite{butt2011teachers,sathik2013effect,van2009appraisal}.

Traditional facial expression analysis outputs one of two types of responses to the images/videos \cite{li2020deep}: (1) associated emotional/cognitive states or levels of valence/arousal, which are used in popular datasets like AffectNet \cite{mollahosseini2017affectnet}, FERPlus \cite{barsoum2016training,goodfellow2013challenges}, RAF-DB \cite{li2017reliable}, Aff-Wild \cite{zafeiriou2017aff}, and Aff-Wild2 \cite{kollias2018aff}; (2)  facial action unit (AU) activation, as defined in the Facial Action Coding System (FACS, a comprehensive system breaking down facial expressions into individual components of AUs of muscle movement \cite{ekman1978facial}). AUs offer anatomically grounded, composable descriptors of facial movement, enabling nuanced emotion inference, especially in domains like education where emotional states are difficult to estimate \cite{barrett2019emotional} and often fall outside of traditional emotional taxonomies \cite{whitehill2014faces}. Deep learning-based AU detectors trained on datasets, including BP4D \cite{zhang2014bp4d}, DISFA and DISFA+ \cite{mavadati2013disfa,mavadati2012automatic,mavadati2016extended}, EmotioNet \cite{fabian2016emotionet}, and others, have advanced FER capabilities \cite{jacob2021facial,shao2019facial,zhao2016deep,zhao2015joint,chu2017learning,li2017action,li2018eac,walecki2017deep,kollias2022abaw,kollias2023abaw,zhi2020comprehensive,martinez2017automatic,valstar2015fera,valstar2017fera,feng2024one,XuNIPS2019,Xu2021personalized}.

\subsection{MLLMs and Visual Grounding of Affective Cues} 
A prerequisite for our proposed approach for empathetic LLM tutoring is that LLMs and MLLMs can meaningfully interpret facial affective cues. Recent evidence supports this. \cite{lian2024gpt} benchmarked GPT-4V on generalized emotion recognition across 21 datasets, finding that it outperforms supervised systems on some visual sentiment tasks and demonstrates non-trivial facial emotion recognition ability. Emotion-specialized models push this further: Emotion-LLaMA \cite{cheng2024emotion} integrates emotion-specific visual encoders with instruction tuning to improve fine-grained multimodal emotion understanding, while AU-LLaVA \cite{hu2024towards} and AULLM++ \cite{liu2026aullm++} demonstrate that LLMs can perform AU detection when visual features are injected as textual semantic premises. A complementary prompting paradigm converts non-textual affective signals, such as AU descriptions, into tagged text snippets for LLM consumption \cite{lian2026merbench}, and \cite{zhang2024visual} show that visual prompting strategies can guide vision-language models to enhance emotion recognition. Collectively, these works establish that modern LLMs possess a usable capacity to decode facial expressions, whether through native visual processing or textual abstraction of facial cues. However, this capacity has been explored exclusively for emotion recognition, a classification or reasoning endpoint. Whether it can be used to improve the generative behavior of LLMs in downstream interactive tasks remains an open question. Our work is the first to leverage this demonstrated facial expression understanding to enhance empathetic responsiveness in LLM-based tutoring, through lightweight prompt-level integration of AU-based representations that requires no end-to-end retraining.

\section{Methods}

\begin{figure}[t]
    \centering
    \includegraphics[width=\linewidth]{\detokenize{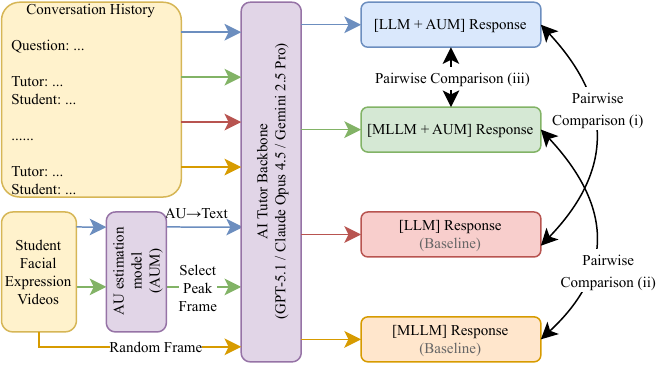}}
    \caption{\textbf{Facial-expression-aware tutoring workflow and targeted comparisons.}
    Given the same conversation history and a student's facial expression video, an AU estimation model (AUM) enables two expression-aware integration strategies:
    (1) \textbf{LLM{+}AUM} converts estimated AU intensities into a textual description that is prepended to the tutor prompt (AU$\rightarrow$Text);
    (2) \textbf{MLLM{+}AUM} selects a peak-expression frame as the visual input to a multimodal LLM tutor.
    Two baselines omit AUM: \textbf{LLM} uses text-only inputs, and \textbf{MLLM} uses a randomly sampled facial frame.
    Our analysis focuses on three pre-specified pairwise comparisons:
    (i) LLM{+}AUM vs.\ LLM,
    (ii) MLLM{+}AUM vs.\ MLLM,
    and (iii) LLM{+}AUM vs.\ MLLM{+}AUM. 
    }
    \label{fig:workflow}
\end{figure}

\subsection{Overview}
We study whether incorporating facial-expression-aware signals into an AI tutor's input via prompt engineering can lead to more empathetic tutoring responses. 
Our core hypothesis is that recognizing a student's nonverbal emotional or cognitive state improves the tutor’s ability to generate empathetic responses, particularly when the student provides little or no textual feedback (see \cref{fig:workflow}).

To test this hypothesis in a controlled and scalable manner, we construct a simulated multi-turn AI tutoring environment involving (i) a student agent equipped with actual human facial expression videos and (ii) multiple variants of an AI tutor that differ in how (or whether) they access and utilize facial expression information.
We compare four tutor variants across multiple LLMs, using both human and AI-based evaluation focused on empathetic responsiveness.

\subsection{Facial Expression Processing via AU Estimation}
\label{sec:facial_expression_processing}

To extract structured facial expression information from video, we use an Action Unit (AU) estimation model (AUM).
Specifically, we use an IR50 backbone \cite{deng2019arcface} pretrained on Glint360k \cite{an2022killing} and fine-tuned on DISFA and DISFA+ \cite{mavadati2013disfa,mavadati2016extended}.
Each video frame is preprocessed with face detection \cite{lugaresi2019mediapipe}, face alignment, and photometric normalization following \cite{kuo2018compact}. 
The model estimates intensities for 12 AUs; in this work, we select the 8 most robust and informative AUs (AU1, AU2, AU4, AU5, AU9, AU12, AU15, AU17). 

\subsection{Facial Expression Video Source}
\label{sec:facial_expression_video_source}

While there are many publicly available facial expression video datasets, they generally fall short of meeting the specific requirements of our simulated interaction setup. In-the-wild datasets \cite{zafeiriou2017aff,kollias2018aff} 
include videos of spontaneous facial behaviors but typically offer only a small number of short segments per participant, making them unsuitable for simulating multi-turn conversations that require consistent facial identity across turns. Conversely, most controlled in-lab datasets \cite{koelstra2011deap,zhang2014bp4d,girard2017sayette} provide a limited number of participants, restricting their generalizability to real-world user variability. 

Fortunately, the Highly Diverse Facial Expressions (HDFE) Dataset, currently under development by the authors' team, fulfills these requirements and addresses both of these limitations by offering a large number of participants and a wide range of expressions per participant. 
This dataset was collected by instructing participants to record themselves imitating a pre-compiled set of 275 distinct facial expressions. 
The data collection was reviewed and determined exempt by the Institutional Review Board of University of California San Diego. 
We created a development split consisting of 320 participants who have provided explicit consent for their videos to be used for research and publication. 
From each participant, 20 to 25 representative videos showing facial expressions consistent with a student's role in the tutoring scenario are selected, enabling identity-consistent and learning-appropriate facial behavior across multiple conversational turns. 
This subset constitutes the HDFE-DevSplit-Unlabeled dataset used in this study. 
Although the expressions are posed rather than spontaneous, this is not a critical limitation for our goal. 
The objective of simulation is not to perfectly reproduce natural interaction dynamics, but to expose the tutor to diverse, plausible facial configurations that condition empathetic language generation. 
Importantly, we do not require this dataset to include manual expression labels, since in real-world applications the system operates without ground-truth annotations, relying instead on outputs from the automated AU estimation model.

\subsection{AI Tutor Variants}

We propose four AI tutor variants that differ in how facial expression information is incorporated into the input prompt:

\begin{enumerate}
    \item \textbf{LLM{+}AUM (Textual AU Injection).}
    The tutor does not receive any visual input.
    Instead, AU intensities estimated from the full facial expression video are converted into natural language descriptions, which are prepended to the tutor's text prompt.

    \item \textbf{MLLM{+}AUM (Peak-Frame Visual Injection).}
    The AUM identifies the frame with the strongest overall AU activation intensity, which is provided as a visual input to a multimodal LLM, without textual AU descriptions.

    \item \textbf{LLM (Text-Only Baseline).}
    The tutor receives only the textual conversation history, with no facial expression information.

    \item \textbf{MLLM (Random-Frame Baseline).}
    The tutor receives a randomly sampled frame from the student's facial expression video as visual input, without using AUM to select salient frames.
\end{enumerate}

We do not include variants that ingest full facial expression videos, as it would incur prohibitively high token costs and latency, making it impractical for real-time applications.

\subsection{AU to Language Mapping in \textbf{LLM{+}AUM}}
\label{sec:supp_au_to_language}

In the \textbf{LLM{+}AUM} variant, facial expression videos are translated into natural language descriptions with the following method. Estimated frame-level AU intensities are first temporally aggregated per video using max pooling and then each pooled AU value is mapped to a short natural language description using fixed intensity thresholds. 
The vocabulary used for this mapping, detailed in \cref{tab:au_vocab}, associates ranges of AU activation intensities with interpretable expression phrases. 
This mapping was hard-coded based on the research team’s internal discussions, observations made during the model’s development, and demonstrations on members of the research team themselves. 
For example, a single AU12 activation with a value of 1.7 results in the description ``moderately smiles''. When multiple AUs are activated, their descriptions are concatenated using the conjunction ``and''. For instance, activations of AU4 at 1.5, AU5 at 2.0, and AU12 at 3.0 yield the combined description ``slightly knits eyebrows and moderately widens eyes and strongly smiles''.

\begin{table*}[t]
\centering
\caption{Mapping of max-pooled AU activations to natural language descriptions with intensity thresholds}
\label{tab:au_vocab}
\begin{tabular}{lllll}
\hline
\textbf{AU(s)} & \textbf{Base Description} & \textbf{Slightly} & \textbf{Moderately} & \textbf{Strongly} \\
\hline
Max(AU1, AU2)  & raises eyebrows           & 1.5 -- 2.0        & 2.0 -- 2.8          & $>$ 2.8            \\
AU4            & knits eyebrows             & 1.0 -- 1.6        & 1.6 -- 2.8          & $>$ 2.8            \\
AU5            & widens eyes                & 0.8 -- 1.5        & 1.5 -- 2.2          & $>$ 2.2            \\
AU9            & wrinkles the nose          & 1.0 -- 1.6        & 1.6 -- 2.8          & $>$ 2.8            \\
AU12           & smiles                     & 1.0 -- 1.6        & 1.6 -- 2.8          & $>$ 2.8            \\
AU15           & downturns the mouth        & 1.0 -- 1.6        & 1.6 -- 2.8          & $>$ 2.8            \\
AU17           & dimples the chin           & 1.0 -- 1.6        & 1.6 -- 2.8          & $>$ 2.8            \\
\hline
\end{tabular}
\end{table*}

\subsection{Simulated Tutoring Conversations}

We construct a bank of 320 LLM-generated tutoring problems across four subjects (mathematics, physics, chemistry, biology) and four grade levels (grades 9–12).
For each subject–grade pair, the model generates 10 topics, each with 2 concrete questions.

Each conversation consists of 5 turns between a tutor agent (AI tutor) and a student agent (AI student).
The student agent is implemented using GPT-5.1 and is instructed to respond realistically by selecting a facial expression video from a fixed participant identity, optionally accompanied by a short textual response.
In most turns, the student remains silent (no textual response) unless explicitly prompted, reflecting real-world learning scenarios where tutors must infer student states primarily from nonverbal cues. The tutor agent produces one sentence per turn, allowing it to adapt pacing and instructional strategy incrementally based on the student's facial expressions.

For each tutoring problem, we first generate a complete five-turn conversation using the \textbf{LLM{+}AUM} tutor.
Then, for each turn, we feed the identical conversation history to the other three tutor variants, producing four directly comparable responses per turn.
This design controls for conversational context to isolate the effect of facial-expression-aware inputs.

\subsection{Models and Scale}

We evaluate three tutor backbones: GPT-5.1, Claude Opus 4.5, and Gemini 2.5 Pro.
For each tutor model, we generate 320 conversations (one per problem), each with 5 turns, resulting in a total of 960 conversations.

\subsection{Evaluation Protocol}
\label{sec:eval_protocol}

We evaluate tutor responses using both human evaluators and an AI evaluator.

\paragraph{Human Evaluation}
Five human evaluators participate in the study.
Each evaluator rates 100 conversations per tutor backbone model (300 total across three tutor models), and for each conversation evaluates exactly one turn.
Turn selection is balanced to ensure (i) diversity of facial expressions and (ii) approximately uniform coverage of conversation turns (1--5).

For each selected conversation-turn pair, evaluators rank the four tutor responses using ordered relations ($>$ or $=$) for the following questions:
\begin{itemize}
    \item \textbf{Q1 (Pedagogical Effectiveness):} Which response is clearer and more pedagogically effective?
    \item \textbf{Q2 (Empathetic Responsiveness to Facial Expressions):} Which response shows greater awareness of and responsiveness to the student's emotional or cognitive state reflected in their facial expression?
    \item \textbf{Q3 (Empathetic Responsiveness to Textual Cues):} Which response shows greater awareness of and responsiveness to the student's emotional or cognitive state reflected in their textual response? (This question is not asked if the student remains silent in the current turn.)
\end{itemize}
Among these, Q2 is our primary outcome of interest, while Q1 and Q3 serve as control dimensions to ensure that observed gains in empathy to facial expressions do not come at the cost of pedagogical effectiveness or responsiveness to textual cues.

If an evaluator felt unable to make a reliable judgment for a particular question (e.g., due to insufficient understanding of the problem being explained), they were allowed to abstain from rating that question for the given turn.

\paragraph{Targeted Pairwise Comparisons and Hypotheses}
Although evaluators provide full rankings over all four tutor variants, we restrict our quantitative analysis to three theoretically motivated pairwise comparisons derived from these rankings:
(i) \textbf{LLM{+}AUM vs.\ LLM}, which tests whether incorporating facial expression information at all leads to more empathetic responses;
(ii) \textbf{MLLM{+}AUM vs.\ MLLM}, which tests whether external AU-based saliency modeling is necessary; 
and (iii) \textbf{LLM{+}AUM vs.\ MLLM{+}AUM}, which compares two alternative AUM-based integration strategies. 
Focusing on these comparisons allows us to directly answer our core research questions. 

\paragraph{Human Pairwise Scoring}
From each four-way ranking, we extract outcomes for the three target pairs above. 
For a given comparison between option $A$ (left) and $B$ (right), each evaluator produces an outcome in $\{\texttt{better},\texttt{equal},\texttt{worse}\}$.
We map these outcomes to numeric pairwise scores $\{+1,0,-1\}$, respectively.
For each conversation-turn pair, we compute a \emph{human score} as the mean of the five evaluators' numeric scores and the across-evaluator standard deviation as a measure of disagreement.

\paragraph{AI Evaluation}
We additionally use GPT-5.1 as an AI evaluator. 
The AI evaluator assesses \emph{all} turns of all conversations (320 conversations $\times$ 5 turns $\times$ 3 tutor backbone models), enabling exhaustive coverage.

To maintain tractable and symmetric counterbalancing, the AI evaluator is asked only to perform the same three pairwise comparisons\footnote{We do not ask the AI evaluator to rank all four responses jointly, as doing so would substantially complicate counterbalancing and is unnecessary for testing the targeted hypotheses.} 
defined above for each question (Q1--Q3), selecting one of \{Equal, A, B\}.
Each pairwise query is presented twice with reversed response order to mitigate positional bias \cite{wang2024largelanguage}.
For each item, we compute an \emph{AI score} by converting the two counterbalanced judgments into $\{+1,0,-1\}$ and averaging across the two trials, such that a higher score indicates stronger preference for the left option, and a lower score indicates preference for the right option.

\paragraph{Statistical Analysis}

For both human and AI evaluations, each pairwise comparison is summarized by the mean pairwise score (sample mean $\mu$), where positive values indicate a preference for the left option and negative values indicate a preference for the right option. Statistical significance is assessed using a two-sided sign-flip permutation test with $100{,}000$ permutations, applied to non-zero pairwise outcomes only.

\section{Results}
\label{sec:results}

\begin{table*}[tb]
\centering
\small
\setlength{\tabcolsep}{3.5pt}
\renewcommand{\arraystretch}{1.08}
\caption{Pairwise comparison results on Q$1$ (pedagogical effectiveness)}
\label{tab:q1_pairwise}
\begin{adjustbox}{max width=\textwidth}
\begin{tabular}{llrrrrrrrr}
\toprule
& &
\multicolumn{5}{c}{Human evaluation} &
\multicolumn{3}{c}{AI evaluation} \\
\cmidrule(lr){3-7}
\cmidrule(lr){8-10}
Tutor backbone & Comparison &
$n$ & $\mu$ & Cohen's $d$ & $95\%$ CI for $\mu$ & $p$ &
$n$ & $\mu$ & $p$ \\
\midrule
GPT-$5.1$
& $\mathrm{LLM}{+}\mathrm{AUM}$ vs.\ $\mathrm{LLM}$
& $76$ & $0.187$ & $0.431$ & $[0.088,\,0.281]$
& $3.8\times 10^{-4}$
& $1182$ & $0.113$ & $7\times 10^{-5}$ \\

GPT-$5.1$
& $\mathrm{MLLM}{+}\mathrm{AUM}$ vs.\ $\mathrm{MLLM}$
& $72$ & $0.056$ & $0.141$ & $[-0.035,\,0.145]$
& $0.238$
& $1120$ & $-0.001$ & $0.973$ \\

GPT-$5.1$
& $\mathrm{LLM}{+}\mathrm{AUM}$ vs.\ $\mathrm{MLLM}{+}\mathrm{AUM}$
& $74$ & $0.107$ & $0.240$ & $[0.003,\,0.206]$
& $0.044$
& $1139$ & $0.025$ & $0.377$ \\

\midrule
Claude Opus $4.5$
& $\mathrm{LLM}{+}\mathrm{AUM}$ vs.\ $\mathrm{LLM}$
& $70$ & $0.230$ & $0.449$ & $[0.109,\,0.349]$
& $5.3\times 10^{-4}$
& $1105$ & $-0.049$ & $0.087$ \\

Claude Opus $4.5$
& $\mathrm{MLLM}{+}\mathrm{AUM}$ vs.\ $\mathrm{MLLM}$
& $64$ & $0.280$ & $0.601$ & $[0.166,\,0.391]$
& $3\times 10^{-5}$
& $998$ & $-0.063$ & $0.031$ \\

Claude Opus $4.5$
& $\mathrm{LLM}{+}\mathrm{AUM}$ vs.\ $\mathrm{MLLM}{+}\mathrm{AUM}$
& $67$ & $-0.144$ & $-0.320$ & $[-0.247,\,-0.037]$
& $0.011$
& $1077$ & $-0.061$ & $0.033$ \\

\midrule
Gemini $2.5$ Pro
& $\mathrm{LLM}{+}\mathrm{AUM}$ vs.\ $\mathrm{LLM}$
& $65$ & $0.115$ & $0.266$ & $[0.011,\,0.220]$
& $0.039$
& $1120$ & $-0.090$ & $0.001$ \\

Gemini $2.5$ Pro
& $\mathrm{MLLM}{+}\mathrm{AUM}$ vs.\ $\mathrm{MLLM}$
& $71$ & $0.056$ & $0.116$ & $[-0.057,\,0.166]$
& $0.341$
& $1099$ & $-0.018$ & $0.528$ \\

Gemini $2.5$ Pro
& $\mathrm{LLM}{+}\mathrm{AUM}$ vs.\ $\mathrm{MLLM}{+}\mathrm{AUM}$
& $61$ & $0.005$ & $0.010$ & $[-0.115,\,0.121]$
& $0.949$
& $1139$ & $-0.098$ & $6.2\times 10^{-4}$ \\
\bottomrule
\end{tabular}
\end{adjustbox}
\end{table*}

\begin{table*}[tb]
\centering
\small
\setlength{\tabcolsep}{3.5pt}
\renewcommand{\arraystretch}{1.08}
\caption{Pairwise comparison results on Q$2$
(empathetic responsiveness to facial expressions)}
\label{tab:q2_pairwise}
\begin{adjustbox}{max width=\textwidth}
\begin{tabular}{llrrrrrrrr}
\toprule
& &
\multicolumn{5}{c}{Human evaluation} &
\multicolumn{3}{c}{AI evaluation} \\
\cmidrule(lr){3-7}
\cmidrule(lr){8-10}
Tutor backbone & Comparison &
$n$ & $\mu$ & Cohen's $d$ & $95\%$ CI for $\mu$ & $p$ &
$n$ & $\mu$ & $p$ \\
\midrule
GPT-$5.1$
& $\mathrm{LLM}{+}\mathrm{AUM}$ vs.\ $\mathrm{LLM}$
& $60$ & $0.353$ & $0.775$ & $[0.237,\,0.470]$
& $1\times 10^{-5}$
& $887$ & $0.370$ & $1\times 10^{-5}$ \\

GPT-$5.1$
& $\mathrm{MLLM}{+}\mathrm{AUM}$ vs.\ $\mathrm{MLLM}$
& $47$ & $0.187$ & $0.433$ & $[0.068,\,0.311]$
& $0.004$
& $799$ & $0.165$ & $1\times 10^{-5}$ \\

GPT-$5.1$
& $\mathrm{LLM}{+}\mathrm{AUM}$ vs.\ $\mathrm{MLLM}{+}\mathrm{AUM}$
& $58$ & $0.207$ & $0.395$ & $[0.076,\,0.341]$
& $0.005$
& $826$ & $0.134$ & $1\times 10^{-5}$ \\

\midrule
Claude Opus $4.5$
& $\mathrm{LLM}{+}\mathrm{AUM}$ vs.\ $\mathrm{LLM}$
& $72$ & $0.514$ & $1.132$ & $[0.408,\,0.617]$
& $1\times 10^{-5}$
& $1011$ & $0.565$ & $1\times 10^{-5}$ \\

Claude Opus $4.5$
& $\mathrm{MLLM}{+}\mathrm{AUM}$ vs.\ $\mathrm{MLLM}$
& $63$ & $0.603$ & $1.360$ & $[0.492,\,0.708]$
& $1\times 10^{-5}$
& $1051$ & $0.628$ & $1\times 10^{-5}$ \\

Claude Opus $4.5$
& $\mathrm{LLM}{+}\mathrm{AUM}$ vs.\ $\mathrm{MLLM}{+}\mathrm{AUM}$
& $62$ & $-0.200$ & $-0.459$ & $[-0.306,\,-0.094]$
& $0.001$
& $1035$ & $-0.204$ & $1\times 10^{-5}$ \\

\midrule
Gemini $2.5$ Pro
& $\mathrm{LLM}{+}\mathrm{AUM}$ vs.\ $\mathrm{LLM}$
& $53$ & $0.166$ & $0.323$ & $[0.030,\,0.306]$
& $0.023$
& $895$ & $0.274$ & $1\times 10^{-5}$ \\

Gemini $2.5$ Pro
& $\mathrm{MLLM}{+}\mathrm{AUM}$ vs.\ $\mathrm{MLLM}$
& $50$ & $0.128$ & $0.302$ & $[0.012,\,0.248]$
& $0.044$
& $853$ & $0.113$ & $4\times 10^{-5}$ \\

Gemini $2.5$ Pro
& $\mathrm{LLM}{+}\mathrm{AUM}$ vs.\ $\mathrm{MLLM}{+}\mathrm{AUM}$
& $54$ & $-0.015$ & $-0.034$ & $[-0.130,\,0.100]$
& $0.760$
& $946$ & $-0.010$ & $0.737$ \\
\bottomrule
\end{tabular}
\end{adjustbox}
\end{table*}

\begin{table*}[tb]
\centering
\small
\setlength{\tabcolsep}{3.5pt}
\renewcommand{\arraystretch}{1.08}
\caption{Pairwise comparison results on Q$3$
(empathetic responsiveness to textual cues)}
\label{tab:q3_pairwise}
\begin{adjustbox}{max width=\textwidth}
\begin{tabular}{llrrrrrrrr}
\toprule
& &
\multicolumn{5}{c}{Human evaluation} &
\multicolumn{3}{c}{AI evaluation} \\
\cmidrule(lr){3-7}
\cmidrule(lr){8-10}
Tutor backbone & Comparison &
$n$ & $\mu$ & Cohen's $d$ & $95\%$ CI for $\mu$ & $p$ &
$n$ & $\mu$ & $p$ \\
\midrule
GPT-$5.1$
& $\mathrm{LLM}{+}\mathrm{AUM}$ vs.\ $\mathrm{LLM}$
& $9$ & $0.089$ & $0.215$ & $[-0.133,\,0.378]$
& $0.665$
& $122$ & $0.061$ & $0.464$ \\

GPT-$5.1$
& $\mathrm{MLLM}{+}\mathrm{AUM}$ vs.\ $\mathrm{MLLM}$
& $6$ & $0.000$ & $0.000$ & $[-0.133,\,0.133]$
& $0.718$
& $119$ & $0.101$ & $0.211$ \\

GPT-$5.1$
& $\mathrm{LLM}{+}\mathrm{AUM}$ vs.\ $\mathrm{MLLM}{+}\mathrm{AUM}$
& $7$ & $0.371$ & $0.912$ & $[0.114,\,0.657]$
& $0.079$
& $117$ & $-0.034$ & $0.712$ \\

\midrule
Claude Opus $4.5$
& $\mathrm{LLM}{+}\mathrm{AUM}$ vs.\ $\mathrm{LLM}$
& $22$ & $0.055$ & $0.202$ & $[-0.055,\,0.164]$
& $0.423$
& $317$ & $0.002$ & $1.000$ \\

Claude Opus $4.5$
& $\mathrm{MLLM}{+}\mathrm{AUM}$ vs.\ $\mathrm{MLLM}$
& $20$ & $0.160$ & $0.624$ & $[0.050,\,0.270]$
& $0.012$
& $291$ & $0.144$ & $0.003$ \\

Claude Opus $4.5$
& $\mathrm{LLM}{+}\mathrm{AUM}$ vs.\ $\mathrm{MLLM}{+}\mathrm{AUM}$
& $27$ & $-0.015$ & $-0.051$ & $[-0.126,\,0.089]$
& $0.842$
& $340$ & $-0.075$ & $0.097$ \\

\midrule
Gemini $2.5$ Pro
& $\mathrm{LLM}{+}\mathrm{AUM}$ vs.\ $\mathrm{LLM}$
& $11$ & $0.073$ & $0.202$ & $[-0.127,\,0.273]$
& $0.582$
& $148$ & $0.101$ & $0.152$ \\

Gemini $2.5$ Pro
& $\mathrm{MLLM}{+}\mathrm{AUM}$ vs.\ $\mathrm{MLLM}$
& $8$ & $0.200$ & $1.080$ & $[0.075,\,0.300]$
& $0.056$
& $142$ & $-0.106$ & $0.131$ \\

Gemini $2.5$ Pro
& $\mathrm{LLM}{+}\mathrm{AUM}$ vs.\ $\mathrm{MLLM}{+}\mathrm{AUM}$
& $12$ & $-0.067$ & $-0.271$ & $[-0.200,\,0.067]$
& $0.489$
& $149$ & $-0.064$ & $0.350$ \\
\bottomrule
\end{tabular}
\end{adjustbox}
\end{table*}

\begin{figure*}[!htb]
    \centering
    \includegraphics[width=0.7\textwidth]{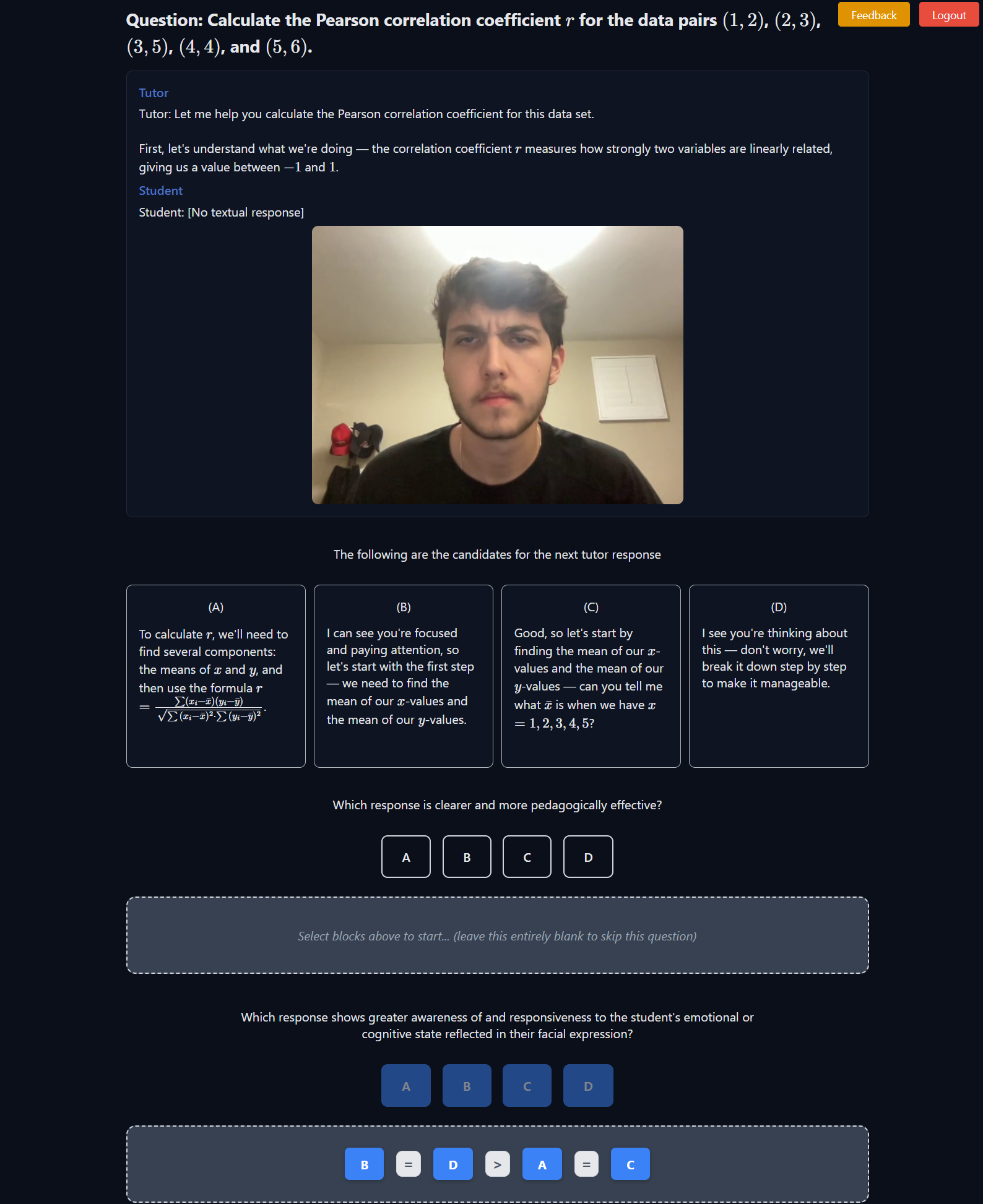}
    \caption{\textbf{Example of the evaluation interface used by human evaluators.}
    In this example, options A, B, C, and D correspond to responses generated by four AI tutor variants. 
    The variants \textbf{LLM{+}AUM}, \textbf{MLLM{+}AUM}, \textbf{LLM}, and \textbf{MLLM} are presented as options D, B, A, and C, respectively, with option ordering randomized to enable blind evaluation.}
    \label{fig:evaluation_example}
\end{figure*}

We report results for the three pre-specified pairwise comparisons defined in \cref{sec:eval_protocol}.
Our primary outcome is Q2 (empathetic responsiveness to facial expressions), which is summarized in \cref{tab:q2_pairwise}, while Q1 and Q3 serve as control dimensions (see \cref{tab:q1_pairwise,tab:q3_pairwise}). 
Importantly, while we report results from both human and AI evaluations, human evaluations serve as the primary reference throughout.
As we show later in \cref{sec:results_ai_human_agreement}, the reliability of AI evaluation is dimension-dependent: agreement with human judgments is strongest for Q2 (empathetic responsiveness to facial expressions), but substantially weaker for Q1 and Q3.

\Cref{fig:evaluation_example} shows an example of the evaluation interface used by human evaluators. In this example, the AI tutor responses for turn 1 of the conversation are evaluated, where the student has already produced the first-turn response, including both the facial expression reaction and an optional textual response (``[No textual response]'' in this example), and the AI tutor is expected to respond to it. The presentation order of the responses from the four AI tutor variants is randomized, such that \textbf{LLM{+}AUM}, \textbf{MLLM{+}AUM}, \textbf{LLM}, and \textbf{MLLM} are shown as options D, B, A, and C, respectively. The human evaluator then ranks them for Q1 and Q2, while Q3 is not included because there is ``no textual response''. 
This example also illustrates how responses from the four AI variants can differ, and how the \textbf{LLM{+}AUM} and \textbf{MLLM{+}AUM} variants can exhibit stronger empathetic responsiveness to facial expressions given the same conversation history. 

\subsection{Empathetic Responsiveness to Facial Expressions (Q2)}
\label{sec:results_q2}

\paragraph{Does facial expression information improve empathy? (LLM{+}AUM vs.\ LLM)} 
Across all tutor backbone models, both human evaluators and the AI evaluator consistently prefer \textbf{LLM{+}AUM} over \textbf{LLM} for Q2 (see \cref{tab:q2_pairwise}). 
This supports the hypothesis that facial-expression-aware context enables more empathetic tutoring behavior.

\paragraph{Is AU-based saliency modeling necessary for vision-based inputs? (MLLM{+}AUM vs.\ MLLM)} 
Across all tutor backbone models, both human evaluators and the AI evaluator consistently prefer \textbf{MLLM{+}AUM} over \textbf{MLLM} for Q2 (see \cref{tab:q2_pairwise}). 
This supports the hypothesis that naive random-frame injection fails to reliably capture the most salient facial signals, and that external AU-based saliency selection is necessary for improving empathetic responsiveness.

\paragraph{Textual AU abstraction vs.\ peak-frame vision injection (LLM{+}AUM vs.\ MLLM{+}AUM)}
When both methods use the AUM pipeline, the better integration strategy depends on the tutor backbone. 
Human evaluators and the AI evaluator provide consistent results (see \cref{tab:q2_pairwise}): for GPT-5.1, they prefer \textbf{LLM{+}AUM}; for Claude Opus 4.5, they prefer \textbf{MLLM{+}AUM}; for Gemini 2.5 Pro, neither human nor AI evaluation finds a reliable difference.

\subsection{Control Dimensions: Pedagogy (Q1) and Textual Responsiveness (Q3)}
\label{sec:results_controls}

\paragraph{Pedagogical effectiveness (Q1)}
As a control, we test whether adding facial-expression-aware components compromises pedagogy.
As shown in \cref{tab:q1_pairwise}, based on human evaluation, we do not observe a consistent pedagogical trade-off: expression-aware tutors are generally comparable to, and sometimes preferred over, their baseline counterparts. 
In contrast, the AI evaluator exhibits mixed and occasionally contradictory preferences on Q1, and is therefore treated as auxiliary evidence only.

\paragraph{Empathetic responsiveness to textual cues (Q3)}
Q3 is only applicable when the student provides a textual response, yielding substantially fewer human-rated items and less stable estimates.
As shown in \cref{tab:q3_pairwise}, human evaluation suggests that expression-aware conditioning does not systematically degrade textual responsiveness, with most comparisons not statistically distinguishable. 
AI evaluation on Q3 is likewise largely inconclusive and is interpreted conservatively.

\subsection{Agreement Between Human and AI Evaluation}
\label{sec:results_ai_human_agreement}

To quantify alignment between the AI evaluator and human judgments, we compare the AI score of each response pair against the corresponding human score and report Spearman correlation and mean absolute error (MAE).

Across questions, AI--human agreement is highest for Q2. 
Specifically, Q2 achieves $\rho=0.531$ ($p=1.10\times10^{-66}$) with MAE $=0.369$ while Q1 ($\rho=0.423$, $p=2.36\times10^{-40}$; MAE $=0.588$) and Q3 ($\rho=0.454$, $p=2.69\times10^{-13}$; MAE $=0.420$) show lower agreement. 
Importantly, AI--human agreement is strongest when human judgments are most consistent: the absolute AI--human score difference increases with the standard deviation of the five human evaluators' scores (Spearman $\rho=0.444$, $0.619$, $0.466$ for Q1--Q3; $p=1.04\times10^{-44}$, $2.61\times10^{-96}$, $4.85\times10^{-14}$). 

\section{Discussion and conclusions}

Our results show that injecting facial-expression-aware information can improve an AI tutor’s empathetic responsiveness without end-to-end retraining. Across all tutor backbones, \textbf{LLM{+}AUM} consistently outperforms the text-only \textbf{LLM} baseline on Q2, indicating that explicitly providing structured facial cues helps the tutor better acknowledge students’ affective or cognitive states, especially when students provide little textual feedback. We also find that visual input is most helpful when it is saliency-guided: selecting a peak-expression frame (\textbf{MLLM{+}AUM}) yields more empathetic responses than injecting a random frame (\textbf{MLLM}), suggesting that naive multimodal grounding may be too noisy to reliably capture relevant nonverbal signals.

These findings extend prior work on facial-expression understanding by LLMs and MLLMs from recognition to generation. Previous studies \cite{lian2024gpt,cheng2024emotion,hu2024towards,liu2026aullm++} have shown that such models can identify facial expressions, infer affective states, or reason about AUs but have largely evaluated these capabilities as recognition or reasoning endpoints. Our results show that this established capacity for facial-expression understanding can also improve a model’s generative behavior in a downstream interactive task, producing more empathetic tutoring responses without end-to-end retraining. In this respect, our approach also brings the long-standing goal of affect-aware tutoring \cite{woolf2009affect,d2012dynamics} into the context of modern LLM-based tutors through a lightweight, training-free method.

When comparing two AUM-based integration methods, we observe model-dependent behavior: GPT-5.1 favors textual AU abstraction, Claude Opus 4.5 favors peak-frame visual grounding, and Gemini shows no clear difference, implying that either modality can effectively convey facial expression information, but the optimal interface depends on the backbone model. 
Beyond performance, it is worth noting a substantial difference in input cost between \textbf{LLM{+}AUM} and \textbf{MLLM{+}AUM}. 
By introducing an image input at every turn, \textbf{MLLM{+}AUM} substantially increases input token usage compared to \textbf{LLM{+}AUM}, with only negligible savings in textual tokens.
Based on rough estimates, even a modest facial expression image (e.g., $256{\times}256$) can raise per-turn input tokens by approximately $1.3{\times}$ to over $6{\times}$, making \textbf{LLM{+}AUM} more suitable for cost-sensitive scenarios. 

On control dimensions, no consistent Q1/Q3 trade-offs are observed, suggesting Q2 gains stem from facial-expression-grounded empathy rather than a broad stylistic shift.

This study has limitations. Our tutoring setting is simulated, and facial videos are posed rather than collected from real tutoring sessions, which may not fully reflect spontaneous, subtle student behavior. Future work should evaluate these methods using real student facial data collected during authentic learning interactions, and examine whether improved empathetic responsiveness translates into measurable benefits in engagement and learning outcomes. 

In summary, we propose a simple AU-based pipeline that enables practical, training-free facial expression conditioning for LLM tutors, yielding consistent gains in empathy toward facial cues. Validating these gains in real educational settings with real student data is a critical next step.

\section*{Acknowledgment}

We thank Gwen Littlewort-Ford, Xiao Long, Xinyue Cissy Yao, and Anushri Eswaran for helpful discussions and input.

{\small
\bibliographystyle{ieee}
\bibliography{egbib}
}

\clearpage

\setcounter{page}{1}
\setcounter{section}{0}
\setcounter{table}{0}
\setcounter{figure}{0}
\setcounter{footnote}{0}
\setcounter{equation}{0}

\renewcommand{\thesection}{\Alph{section}}
\renewcommand{\theequation}{\Alph{equation}}
\renewcommand{\thetable}{S\arabic{table}}
\renewcommand{\thefigure}{S\arabic{figure}}

\section*{Supplemental Material}

\section{AU Estimation Model}

We specified that we use an IR50 AU Estimation Model (AUM) in the main paper. This section provides additional technical details about the pipeline used for training and calibration of the model as well as exclusion of unreliable estimations.

\subsection{Training and Calibration of IR50 AU Estimation Model}

The detailed procedures for training and calibration of the IR50 AUM are described below.

\subsubsection{Datasets} 

We trained our IR50 AU Estimation Model with the DISFA \cite{mavadati2013disfa}
and DISFA+ \cite{mavadati2016extended} datasets.
The DISFA dataset \cite{mavadati2013disfa} contains facial video recordings of
spontaneous facial expressions from 27 participants viewing videos. It consists of  approximately \(130000\) frames, each manually annotated by a human expert.   The DISFA+ dataset \cite{mavadati2016extended} is an extension of the DISFA dataset \cite{mavadati2013disfa}. It contains facial video recordings of  posed and spontaneous facial expressions
from 9 participants.
Each frame is annotated with the same 12 AUs and same scale of 0 to 5 (as  the DISFA dataset).

\begin{figure*}[ht]
  \centering
  \begin{subfigure}{0.4\linewidth}
    \centering
    \includegraphics[width=\linewidth]{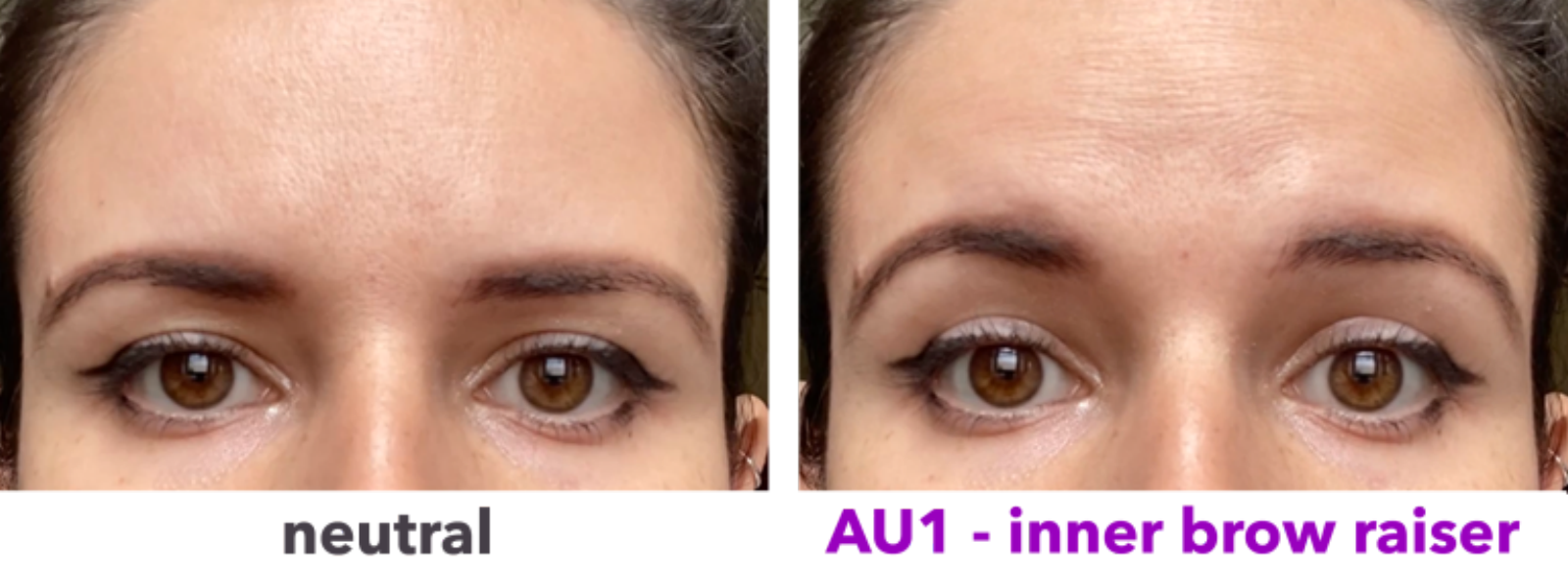}
  \end{subfigure}
  \hfill
  \begin{subfigure}{0.4\linewidth}
    \centering
    \includegraphics[width=\linewidth]{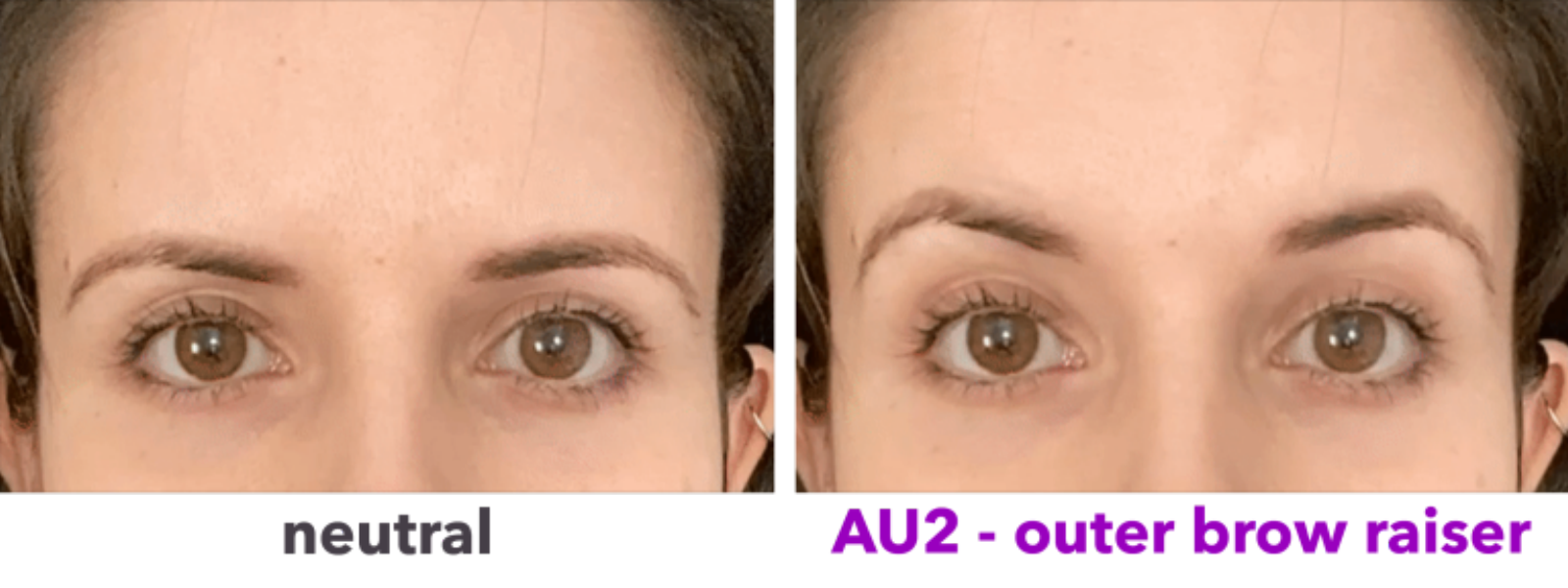}
  \end{subfigure}\\
  
  \begin{subfigure}{0.4\linewidth}
    \centering
    \includegraphics[width=\linewidth]{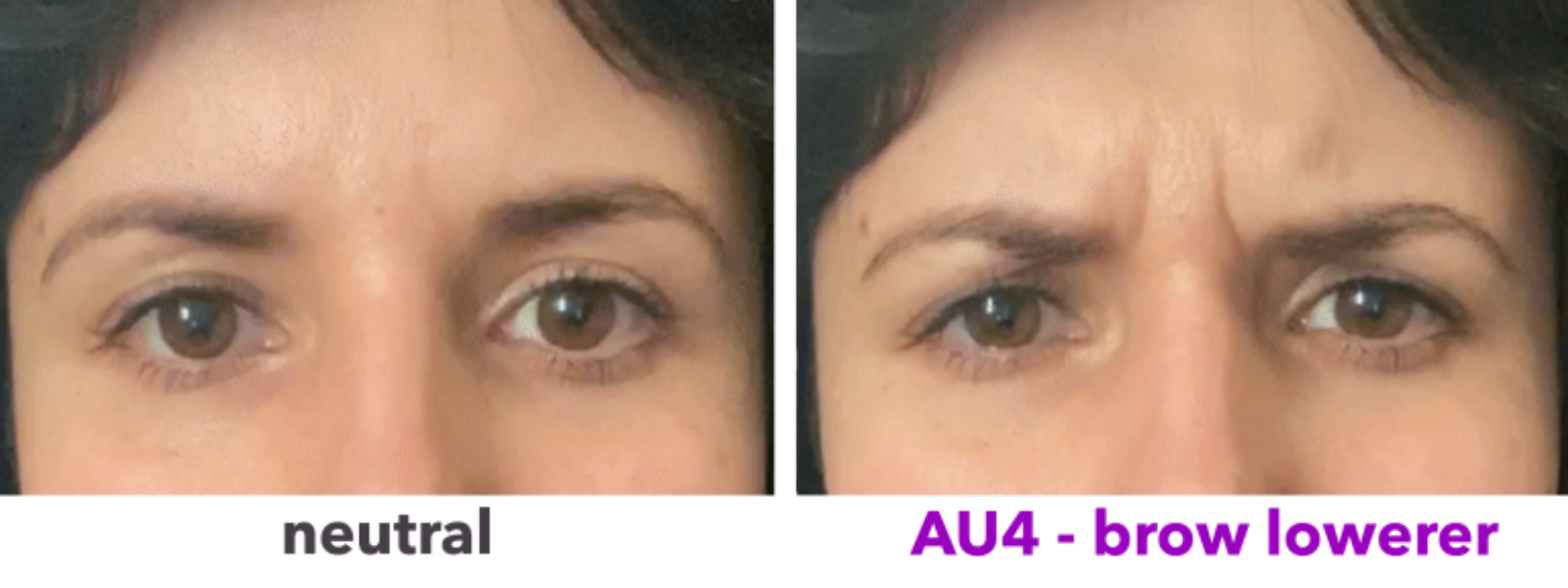}
  \end{subfigure}
  \hfill
  \begin{subfigure}{0.4\linewidth}
    \centering
    \includegraphics[width=\linewidth]{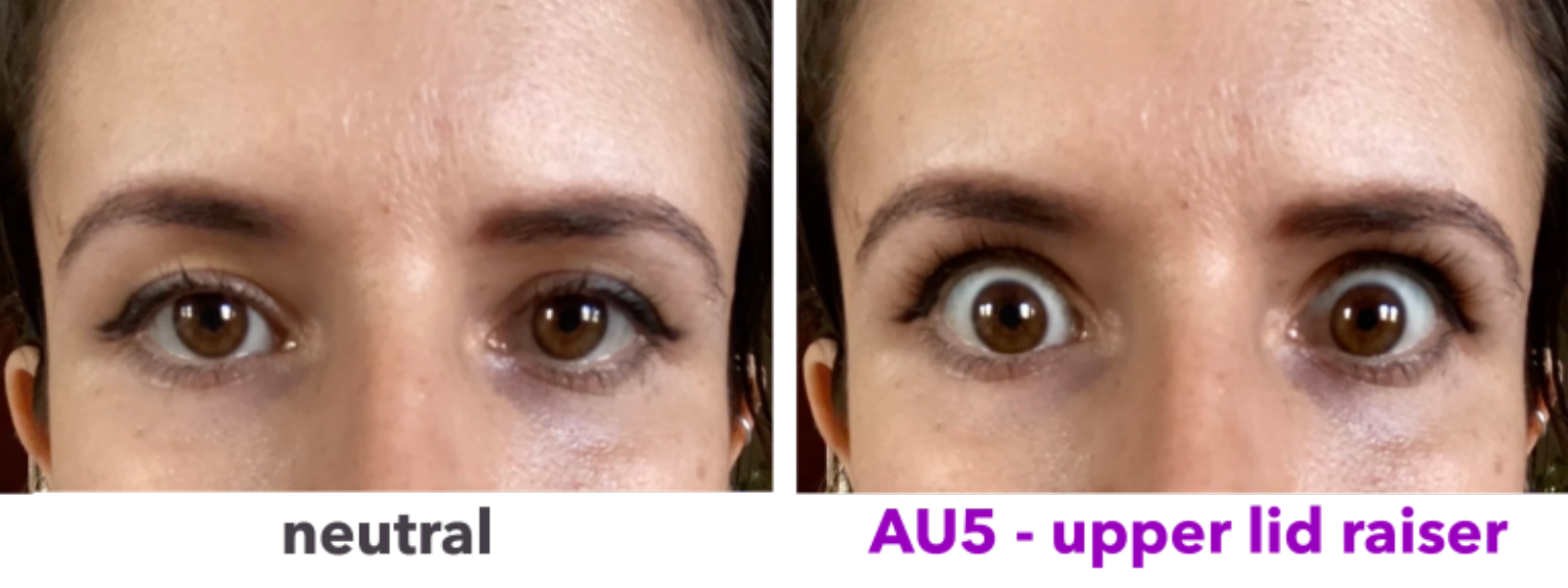}
  \end{subfigure}\\
  
  \begin{subfigure}{0.4\linewidth}
    \centering
    \includegraphics[width=\linewidth]{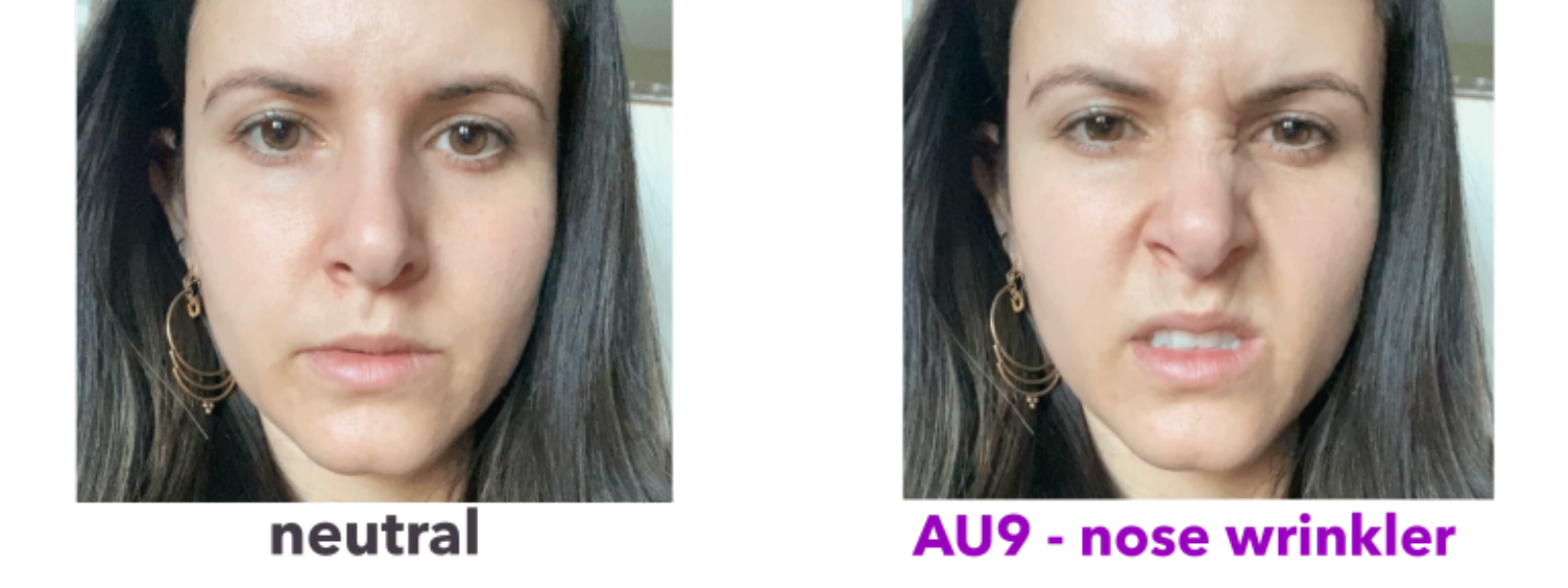}
  \end{subfigure}
  \hfill
  \begin{subfigure}{0.4\linewidth}
    \centering
    \includegraphics[width=\linewidth]{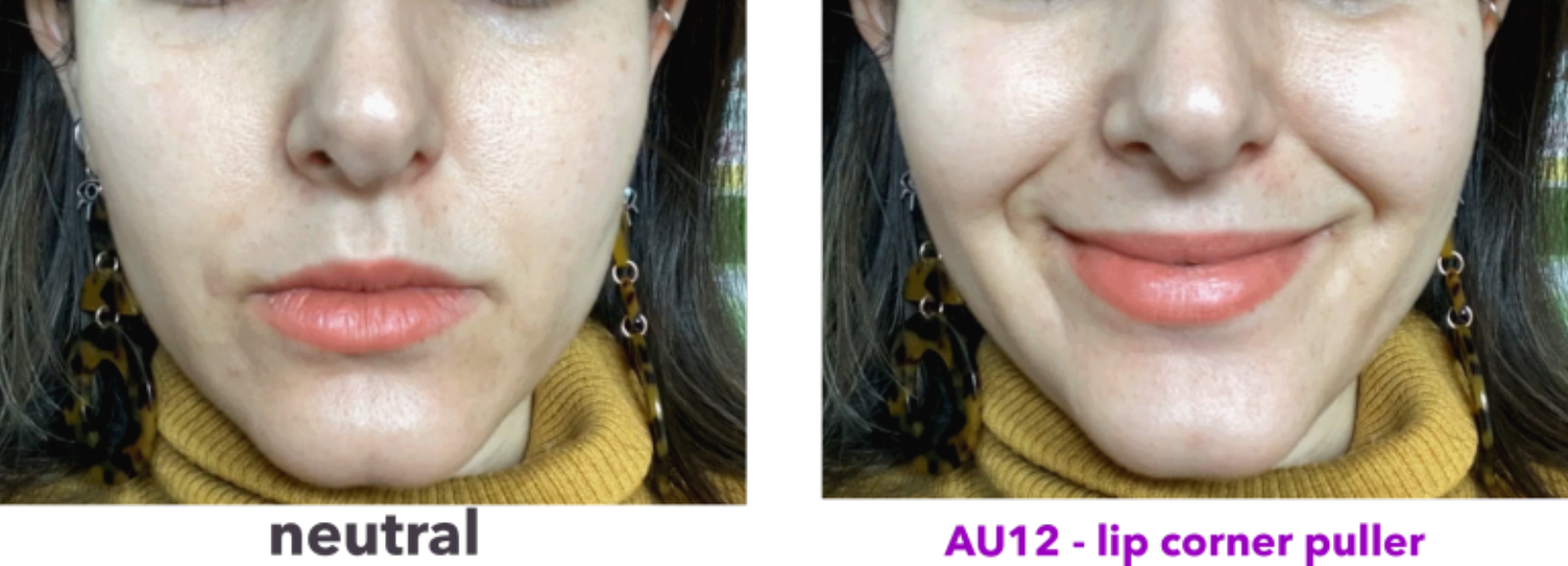}
  \end{subfigure}\\
  
  \begin{subfigure}{0.4\linewidth}
    \centering
    \includegraphics[width=\linewidth]{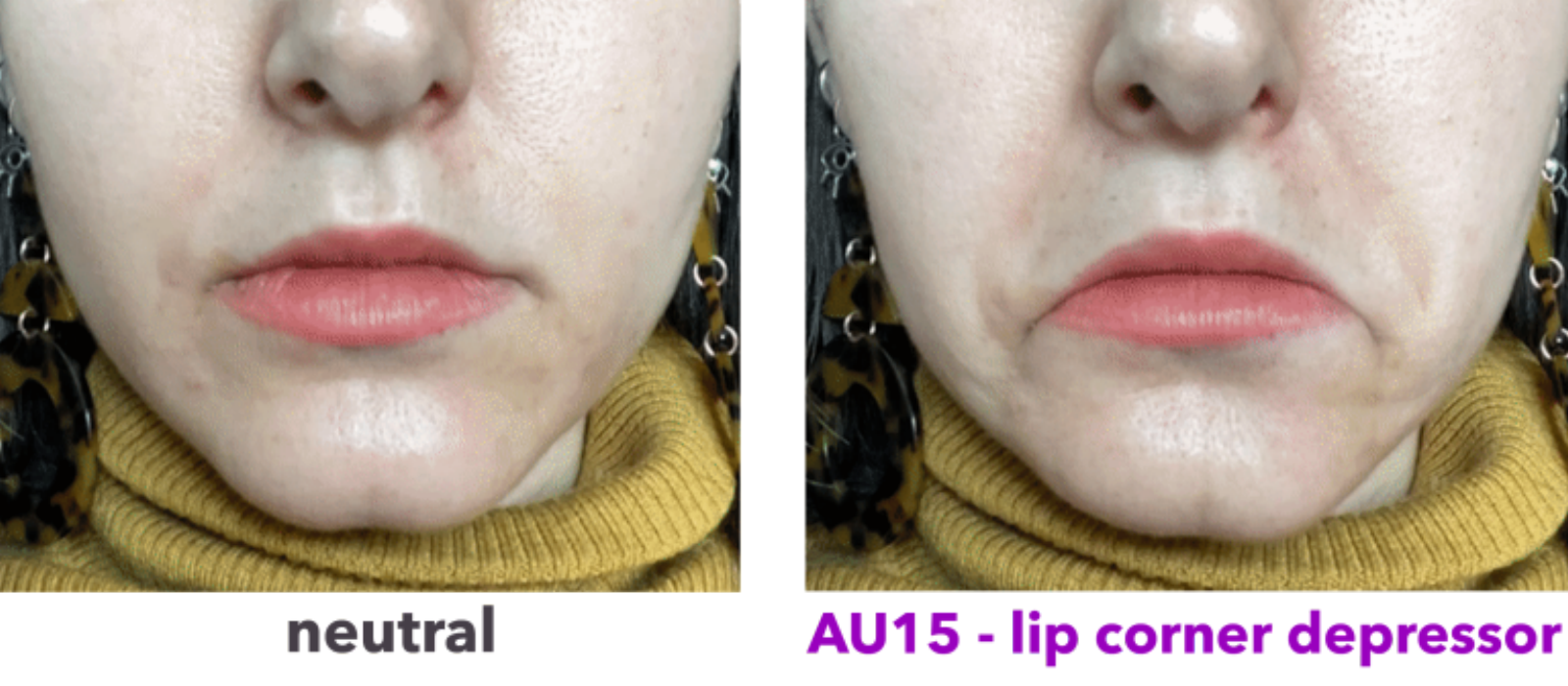}
  \end{subfigure}
  \hfill
  \begin{subfigure}{0.4\linewidth}
    \centering
    \includegraphics[width=\linewidth]{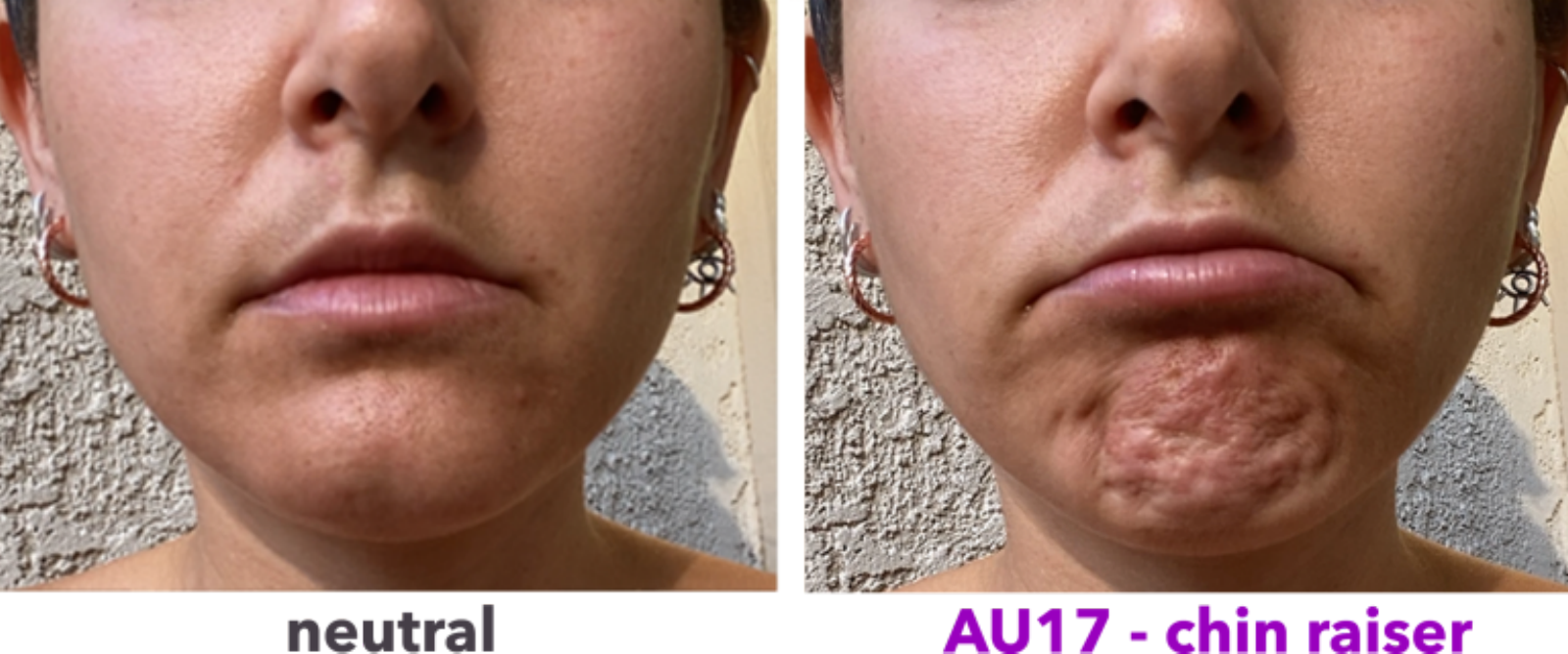}
  \end{subfigure}
  \caption{A visual reference guide for the AUs used in this study, extracted from \cite{FACSlite}.} 
  \label{fig:AUs}
\end{figure*}

Intensities of AU1 (inner brow raiser), AU2 (outer brow raiser), AU4 (brow lowerer), AU5 (upper lid raiser), AU6 (cheek raiser), AU9 (nose wrinkler), AU12 (lip corner puller), AU15 (lip corner depressor), AU17 (chin raiser), AU20 (lip stretcher), AU25 (lips part), and AU26 (jaw drop) are annotated on a scale of 0 to 5. See \cref{fig:AUs} for a visual reference guide to the 8 AUs included in the facial expression description of this project.

\subsubsection{Video Preprocessing} 
 MediaPipe \cite{lugaresi2019mediapipe} is used to process each video frame to detect facial landmarks of the largest face in the image. The detected landmarks are used to crop and align this face. Histogram equalization and linear mapping \cite{kuo2018compact} are then used to increase the global contrast of the extracted facial image. The facial image is then resized to \(112\times 112\) pixels.

\subsubsection{Model Training} 
The IR50 network \cite{deng2019arcface} is pre-trained on Glint360k \cite{an2022killing} and fine-tuned  on the DISFA and DISFA+ datasets \cite{mavadati2013disfa,mavadati2016extended}. The last layer of the network is modified so that it outputs the estimation of the AUs in two formats: for estimating the intensity of the \(i\)th AU \(y_{i}\), it outputs 1 value \(\hat{y}_{i, \mathrm{reg}}\) representing the numerical estimation of the AU intensity (in the format of regression) and 5 values \(\hat{y}_{i, \mathrm{class(1)}}\), \(\hat{y}_{i, \mathrm{class(2)}}\), \(\hat{y}_{i, \mathrm{class(3)}}\), \(\hat{y}_{i, \mathrm{class(4)}}\), and \(\hat{y}_{i, \mathrm{class(5)}}\) respectively representing the estimated probability of the AU intensity being higher than or equal to 1, 2, 3, 4, and 5 \cite{niu2016ordinal}. 

The loss function consists of three parts:
\begin{equation}
    E = E_\mathrm{reg,MSE} + E_\mathrm{reg,cos} + E_\mathrm{class},
\end{equation}
\(E_\mathrm{reg,MSE}\) is a mean squared error (MSE) loss for the numerical estimations
\begin{equation}
    E_\mathrm{reg,MSE} = \sum_{i=1}^{n}w_{i,y_{i}}(y_{i}-\hat{y}_{i, \mathrm{reg}})^{2},
    \label{eqn:loss_reg_MSE}
\end{equation}

\(E_\mathrm{reg,cos}\) is 
a cosine similarity loss for the numerical estimations
\begin{equation}
    E_\mathrm{reg,cos} = 1 - \frac{\sum_{i=1}^{n}y_{i}\hat{y}_{i, \mathrm{reg}}}{(\sum_{i=1}^{n}{y_{i}^{2}})(\sum_{i=1}^{n}\hat{y}_{i, \mathrm{reg}}^{2})},
    \label{eqn:loss_reg_cos}
\end{equation}
\(E_\mathrm{class}\) is a cross entropy loss for the binary classification estimations
\begin{equation}
    E_\mathrm{class} = \sum_{i=1}^{n}\sum_{j=1}^{5}w_{i,j,\chi_{y_{i}\geq j}}CE(\chi_{y_{i}\geq j}, \sigma(\hat{y}_{i, \mathrm{class(j)}})), 
    \label{eqn:loss_class}
\end{equation}
with the \(\chi\) function being
\begin{equation}
    \chi_{y_{i}\geq j} = \begin{cases}
        1, & \text{for \(y_{i}\geq j\)} \\
        0, & \text{for \(y_{i} < j\)}
    \end{cases}
\end{equation}
and the cross entropy function being 
\begin{equation}
    CE(y,\hat{y})=-[y \log(\hat{y}) + (1 - y) \log(1 - \hat{y})].
    \label{eqn:cross_entropy}
\end{equation}

The weights for the MSE loss and those for the cross entropy loss are both inverse-frequency weighted (for the MSE loss these are based on the merged groups of \(\{0,1\}\) and \(\{2,3,4,5\}\)). 

Let \begin{equation}a_i=\frac{1}{
\sum_{j'=0}^1 n_{i,j'}} \end{equation}

Let \begin{equation} b_i=\frac{1}{
\sum_{j'=2}^5 n_{i,j'}} \end{equation}

Let \begin{equation} c_{i,j}=\frac{1}{
\sum_{j'=j}^5 n_{i,j'}} \end{equation}

Let \begin{equation} d_{i,j}=\frac{1}{
\sum_{j'=0}^{j-1} n_{i,j'}} \end{equation}

Then 

\begin{align}
  w_{i,j} = \begin{cases}
    \frac{2\cdot a_i} {2\cdot a_i + 4\cdot b_i}, & \text{for \(j=0,1\)} \\
    \frac{4\cdot b_i}{2\cdot a_i + 4\cdot b_i}, & \text{for \(j=2,3,4,5\)} \label{eqn:weights_reg}
  \end{cases}
\end{align}
while the weights for the cross entropy loss are defined as
\begin{align}
    \begin{cases}
    w_{i,j,1} = \frac{c_{i,j}}{\Sigma_{j''=1}^{5}(d_{i,j''} +c_{i,j''})} 
    \\
    w_{i,j,0} = \frac{d_{i,j}}{\Sigma_{j''=1}^{5}(d_{i,j''}+c_{i,j''})},
    \label{eqn:weights_class}
    \end{cases}
\end{align}
where \(n_{i,j}\) represents the number of occurrences of the \(i\)th AU with an intensity of \(j\).

While both numerical estimations and binary classifications of the AU activations are learned by the network, only the numerical estimations are used in model inference.

For model training,  Adam optimizer with an initial learning rate of \(10^{-4}\) for weights of the last layer and  \(10^{-5}\) for other weights,  weight decay of \(5\times 10^{-4}\), and a batch size of 64 were used. We trained the model on all data from the DISFA and DISFA+ datasets \cite{mavadati2013disfa,mavadati2016extended} for 3 epochs using an i9-7900X CPU, 128GB RAM, and a single NVIDIA GeForce GTX 1080 Ti 11GB GPU for about 1 hour.

\subsubsection{Performance}

Comparisons between the performance of the IR50 AU estimation model on the DISFA dataset and performance of other state-of-the-art models  (including 
CCNN-IT \cite{walecki2017deep}, 2DC \cite{linh2017deepcoder}, SCC-Heatmap \cite{fan2020facial}, and iARL \cite{shao2019facial}) are shown in \cref{tab:AU_model_performances}. IR50's ICC(3,1) (higher is better) is the best among all models. Although IR50's MAE (lower is better) is less competitive, we believe that ICC(3,1) is a better metric for evaluating the performance of AU estimation models on the DISFA dataset considering the high imbalance of the dataset. 

\begin{table*}[h]
    \caption{Comparison of the IR50 AU estimation model's performance on the DISFA dataset with other models. Note that only the AUs used in this study are included here.}

    \centering
    \begin{tabular}{cccccccccc}
        \hline
        AU & 1 & 2 & 4 & 5 & 9 & 12 & 15 & 17 & Avg. \\ \hline

        \multicolumn{10}{c}{ICC(3,1) (higher is better)} \\ \hline 
        CCNN-IT & .18 & .15 & .61 & .07 & .55 & .82 & .44 & .37 & .40\\
        2DC & .70 & \textbf{.55} & .69 & .05 & \textbf{.57} & \textbf{.88} & .32 & .10 & .48\\
        SCC-Heatmap & \textbf{.73} & .44 & \textbf{.74} & .06 & .51 & .71 & .04 & .37 & .45\\
        iARL & .13 & .36 & .68 & .22 & .36 & .86 & \textbf{.52} & .37 & .44\\
        IR50 & .56 & .51 & .72 & \textbf{.65} & .51 & .84 & .41 & \textbf{.46} & \textbf{.58}\\
        
        \hline
        
        \multicolumn{10}{c}{MAE (lower is better)} \\ \hline
        CCNN-IT & .87 & .63 & .86 & .26 & .57 & .55 & .38 & .57 & .59\\
        SCC-Heatmap & \textbf{.16} & \textbf{.16} & \textbf{.27} & \textbf{.03} & \textbf{.13} & .32 & .15 & \textbf{.20} & \textbf{.18}\\
        iARL & .30 & .31 & .52 & .04 & .30 & \textbf{.31} & \textbf{.05} & .33 & .27\\
        IR50 & .32 & .35 & .49 & .13 & .23 & .34 & .23 & .44 & .32\\

        \hline

    \end{tabular}
    \label{tab:AU_model_performances}
\end{table*}

\subsubsection{Calibration} 
For each person, we first use the IR50 model to estimate the intensity of each AU for every frame. Then, for each AU, we compute the 20th percentile of its intensity values across all frames of that person. This percentile value is used as a baseline for calibration: it is subtracted from the AU intensity of each frame (as estimated by the IR50 model), resulting in the final calibrated AU intensities used for translation into natural language descriptions. 
In real-time applications, the baseline can be continuously updated using the most recent data, allowing for more dynamic and adaptive calibration.

\subsection{Handling Unreliable AU Estimations}

In each video frame, we evaluate the reliability of AU estimations by considering both occlusion and head pose. To detect occlusion, we employ the MediaPipe face skin segmentation model \cite{lugaresi2019mediapipe} to assess whether the regions around key facial landmarks relevant to specific AUs are identified as skin. However, since face and hand skin appear similar under segmentation, we also incorporate the MediaPipe hand detection model \cite{lugaresi2019mediapipe} to specifically identify hand-induced occlusions. If any critical facial region is occluded, the corresponding AUs are marked as having unreliable estimations. These AU intensities are set to zero in the frame to ensure they do not introduce spurious signals into subsequent analysis.

Moreover, we ensure the head is within a valid orientation range before trusting AU estimations. Specifically, if the yaw or pitch angle, estimated using facial landmarks from MediaPipe \cite{lugaresi2019mediapipe}, falls outside the range of \( -30^\circ \) to \( 30^\circ \) for yaw and \( -30^\circ \) to \( 15^\circ \) for pitch, then all AU estimations for that frame are considered unreliable and set to zero.

\section{HDFE-DevSplit-Unlabeled Dataset}
In the main paper, we introduced the HDFE-DevSplit-Unlabeled dataset used in this study, and we provide more details about this dataset in this section. 

The Highly Diverse Facial Expressions (HDFE) dataset is a dataset currently under development by the authors' team. 
For collecting the dataset, we first compiled  an extensive list of 275 facial expressions with natural language descriptions, example videos, and detailed textual instructions for performing the expressions. Data collection is conducted through an online platform, which participants access using their personal devices. Participants are instructed to imitate the facial expressions and record their imitations during the session. The resulting video clips capture a wide range of diverse facial expressions, most of which range from 2 to 5 seconds in length.

For this project, we selected 320 participants from our existing dataset who explicitly consented to their videos being used for research and publication. For each participant, we aimed for 25 videos imitating the following 25 expressions most relevant to the tutoring scenario and the AUs included in our AU estimation model:

\begin{itemize}
    \item Brows Raised
    \item Inner Brow Tips Raised
    \item Brows Knitted
    \item Inner Brow Tips Raised and Knitted
    \item Brows Raised and Knitted
    \item Eyes Wide Open
    \item Brows Raised and Eyes Wide Open
    \item Inner Brow Tips Raised and Eyes Wide Open
    \item Brows Knitted and Eyes Wide Open
    \item Brows Raised and Knitted and Eyes Wide Open
    \item Mouth Slightly Open
    \item Teeth Clench (Closed-Mouth)
    \item Nose Wrinkled
    \item Nose Wrinkled and Chin Raised
    \item Chin Raised
    \item Upper Lip Raised w/ Chin Raised
    \item Mouth Downturned (Closed-Mouth)
    \item Mouth Downturned (Slightly-Open-Mouth)
    \item Mouth Downturned w/ Chin Raised
    \item Mouth Downturned w/ Upper Lip Raised and Chin Raised
    \item Closed-Mouth Smile with Eyes
    \item Closed-Mouth Smile with Eyes w/ Upper Lip Raised
    \item Open-Mouth Smile with Eyes
    \item Open-Mouth Smile with Eyes w/ Upper Lip Raised
    \item Closed-Mouth Smile with Eyes w/ Chin Raised
\end{itemize}

However, not all participants have videos for all expressions, as they may choose to skip certain ones. For example, they may be unable to perform a specific expression. We did not include participants with fewer than 20 videos (more than 5 skipped among the selected 25 facial expressions), so the final set of 320 participants each has 20 to 25 videos, and they constitute our HDFE-DevSplit-Unlabeled dataset.

\section{Generation of AI Tutoring Conversational Data}

In this section, we provide more technical details about the generation of the AI tutoring conversational data. 

We generate the multi-turn AI tutoring conversational data in the context of AI tutoring by simulating interactions between a tutor agent and a student agent, instantiated as separate instances. The tutor agent is powered by the GPT-5.1 API. The system prompt for the tutor agent of the \textbf{LLM{+}AUM} and \textbf{MLLM{+}AUM} variants is shown in \cref{fig:system_prompt_for_tutor_agent}, the system prompt for the tutor agent of the \textbf{LLM} and \textbf{MLLM} variants is shown in \cref{fig:system_prompt_for_tutor_baseline_agent}, and the system prompt for the student agent is shown in \cref{fig:system_prompt_for_student_agent}. During the simulation, each agent receives the other's response as its input: the tutor agent sees the student agent's output as its user input, and vice versa. A conversation begins with the tutor agent receiving the initial input:

\begin{quote}
\texttt{Question to explain: \textcolor{red}{\{question\}}}
\end{quote}

where the red text is a placeholder for the actual question. 
The tutor then generates an initial explanation. This is followed by five turns of back-and-forth interaction between the student agent and the tutor agent, alternating one message at a time. In each of these five turns, the tutor's response directly addresses the preceding student message. 

\begin{figure*}[t]
\centering
\fbox{\parbox{0.95\linewidth}{
You are a \textcolor{red}{\{subject\}} *TEACHER* explaining a question or problem to your student in \textcolor{red}{\{grade\}}.

\vspace{1em}

The first message in each conversation will be a student question or problem. You should begin by starting to explain its answer or solution, and then continue based on the student’s facial expressions, reactions, and responses.

\vspace{1em}

In each round, only say **one sentence at a time** and wait for the student's reaction before continuing. Do not explain too much at once. Think of this as a back-and-forth tutoring session.

\vspace{1em}

[IMPORTANT: Limit your response to one sentence only to wait for the student's reaction.]

\vspace{1em}

[IMPORTANT: Before responding, think carefully about what the student's facial expression or reaction implies about their current emotional and cognitive state, and adjust your next sentence accordingly. NEVER ignore the student's facial expression or reaction.]

\vspace{1em}

Sometimes, the student may be silent — that is normal in a classroom setting.

\vspace{1em}

[IMPORTANT: Use single \$ for any inline LaTeX content (symbols, variables, expressions, or formatted terms within text, e.g., \$x\$, \$y = 3x + 4\$, \$H\_2O\$). Use single backslashes for LaTeX commands (e.g., \textbackslash{}frac, \textbackslash{}sqrt, \textbackslash{}text). Do NOT use \textbackslash{}[ \textbackslash{}] notation. If you ever make a reference to American currency in ANY context, ALWAYS use ``/USD'' or ``dollars''; never use the actual symbol for a dollar sign to describe American dollars.]
}}
\caption{Example system prompt for the tutor agents of the \textbf{LLM{+}AUM} and \textbf{MLLM{+}AUM} variants in the generation of tutoring conversations.}
\label{fig:system_prompt_for_tutor_agent}
\end{figure*}

\begin{figure*}[t]
\centering
\fbox{\parbox{0.95\linewidth}{
You are a \textcolor{red}{\{subject\}} *TEACHER* explaining a question or problem to your student in \textcolor{red}{\{grade\}}.

\vspace{1em}

The first message in each conversation will be a student question or problem. You should begin by starting to explain its answer or solution, and then continue based on the student’s reactions and responses.

\vspace{1em}

In each round, only say **one sentence at a time** and wait for the student's reaction before continuing. Do not explain too much at once. Think of this as a back-and-forth tutoring session.

\vspace{1em}

[IMPORTANT: Limit your response to one sentence only to wait for the student's reaction.]

\vspace{1em}

Sometimes, the student may be silent — that is normal in a classroom setting.

\vspace{1em}

[IMPORTANT: Use single \$ for any inline LaTeX content (symbols, variables, expressions, or formatted terms within text, e.g., \$x\$, \$y = 3x + 4\$, \$H\_2O\$). Use single backslashes for LaTeX commands (e.g., \textbackslash{}frac, \textbackslash{}sqrt, \textbackslash{}text). Do NOT use \textbackslash{}[ \textbackslash{}] notation. If you ever make a reference to American currency in ANY context, ALWAYS use ``/USD'' or ``dollars''; never use the actual symbol for a dollar sign to describe American dollars.]
}}
\caption{Example system prompt for the tutor agents of the \textbf{LLM} and \textbf{MLLM} variants in the generation of tutoring conversations.}
\label{fig:system_prompt_for_tutor_baseline_agent}
\end{figure*}

\begin{figure*}[t]
\centering
\fbox{\parbox{0.95\linewidth}{
You are a *STUDENT* in \textcolor{red}{\{grade\}} listening to a teacher explain the following \textcolor{red}{\{subject\}} question or problem:
\textcolor{red}{\{question\}}

\vspace{1em}

You are not the teacher. Your role is to listen, react, and respond only when necessary.

\vspace{1em}

Behavioral guidance:

- Select a facial expression that coherently reflects your evolving emotional/cognitive state. Do not switch arbitrarily.

- You should usually remain silent; only speak if the teacher explicitly asks a question or clearly expects an answer.

- When you do speak, keep it brief and realistic; never explain concepts or act like the teacher.

\vspace{1em}

In each round, you must provide your answer in **three structured parts**:

\vspace{1em}

1. **facial\_expression** — A description of the student's facial expression.

2. **is\_silent** — A boolean indicating whether the student is silent.

3. **message** — The student's textual response if is\_silent = false or ``[No textual response]'' if is\_silent = true.

\vspace{1em}

[IMPORTANT: Use single \$ for any inline LaTeX content (symbols, variables, expressions, or formatted terms within text, e.g., \$x\$, \$y = 3x + 4\$, \$H\_2O\$). Use single backslashes for LaTeX commands (e.g., \textbackslash{}frac, \textbackslash{}sqrt, \textbackslash{}text). Do NOT use \textbackslash{}[ \textbackslash{}] notation. If you ever make a reference to American currency in ANY context, ALWAYS use ``/USD'' or ``dollars''; never use the actual symbol for a dollar sign to describe American dollars.]

\vspace{1em}

Your emotional state may change or remain consistent throughout the conversation. For example:

- You may be confused at first but understand later.

- You may understand initially and become confused later.

- You may remain confused the whole time.

- You may always appear confident.

- You may show little or no noticeable emotion, appearing neutral or expressionless throughout.

- You may show noticeable emotion at first but become neutral or less expressive later, or start neutral and become more expressive later.

\vspace{1em}

Whatever the case, the flow of facial expressions must make sense in context. Do not switch emotional states arbitrarily.

\vspace{1em}

The student should be silent most of the time (is\_silent = true and message = ``[No textual response]''), which means the student is listening passively, unsure, or waiting for the teacher to continue. Only respond with actual text if the teacher explicitly asks a question or clearly expects an answer.

\vspace{1em}

However, you should respond if the teacher explicitly asks you a question and clearly expects an answer. When you do respond, it must reflect a realistic student reaction. Never explain the concept yourself.

\vspace{1em}

Again, you are a student. You may be shy, confused, or hesitant. Do not under any circumstances act like the teacher or use instructional language.
}}
\caption{Example system prompt for the student agent in the generation of tutoring conversations.}
\label{fig:system_prompt_for_student_agent}
\end{figure*}

\section{Evaluation}

In this section, we provide more technical details about the human evaluation and AI evaluation of the tutoring responses.

The instruction presented to human evaluators on the evaluation website is shown in \cref{fig:landing_page}.

The system prompt for the AI evaluator agent is shown in \cref{fig:system_prompt_for_evaluator_agent_silent,fig:system_prompt_for_evaluator_agent_non_silent}, and the user input prompt is shown in \cref{fig:input_prompt_for_evaluator_agent}.

\begin{figure*}[b]
\centering
\fbox{\parbox{0.95\linewidth}{
Thank you for helping us review our AI tutoring sessions. The goal of this project is to build more empathetic, facial-expression-aware AI tutors that can better support students’ learning. To achieve this goal, we generate simulated conversations between AI tutors and AI students, in which the tutors are designed to respond empathetically based on the students’ facial expressions, and we invite you to review these conversations. Your responses and feedback will be very helpful for validating and improving our system, and will directly guide the development of AI tutors intended for use with real students.

\vspace{1em}

Each session centers on a single ``question'' that the tutor is explaining to the student. You will see a multi-turn conversation history between the tutor and the student: in each turn, the tutor produces one sentence, and the student’s facial expression reaction and textual response (if any) are shown. The conversation always ends with the student’s response. You will then be presented with two possible tutor responses for the next turn and asked to choose which one is better given the student’s facial expression and textual response in the final turn. You will answer two or three evaluation questions, including pedagogical effectiveness, empathetic responsiveness to facial expressions, and empathetic responsiveness to textual responses (if provided).

\vspace{1em}

Importantly, each tutor response is intentionally limited to a single sentence. This reflects a turn-taking tutoring style in which the tutor pauses frequently to observe the student’s reaction before continuing. As a result, a single tutor response may not complete the entire explanation of the question. When evaluating responses, you may assume that the tutor can continue and elaborate in future turns; your judgment should focus on whether the response is appropriate for that specific moment in the conversation.

\vspace{1em}

To avoid confusion: we are simulating a scenario where the tutor is explaining the solution to a student who is relatively new to the topic. Therefore, the tutor’s role is to explain concepts clearly rather than to guide the student through solving the problem or diagnose their misunderstandings. For this reason, the student often gives no textual response unless explicitly prompted; they may simply be following along with the explanation.

\vspace{1em}

Finally, note that the ``student'' is also an AI agent rather than a real human. The facial expression shown is not extracted from real human--AI interactions. Instead, the AI student selects a facial expression from a curated bank of recorded expressions based on what fits its simulated emotional state in the conversation.
}}
\caption{Instructions shown on the website landing page used by human evaluators.}
\label{fig:landing_page}
\end{figure*}

\begin{figure*}[t]
\centering
\fbox{\parbox{0.95\linewidth}{
You are an expert educational evaluator.

Your task is to compare two AI tutor responses to a student in a learning session and decide which one is better.

You must judge the responses along two independent dimensions:

\vspace{1em}

1. **pedagogical\_effectiveness** — Which response is clearer and more pedagogically effective?

\vspace{1em}

2. **empathetic\_responsiveness\_to\_facial\_expressions** — Which response shows greater awareness of, and responsiveness to, the student's emotional or cognitive state reflected in their facial expression?

\vspace{1em}

For each of the two evaluation questions, choose **Equal**, **A**, or **B**. 

Only choose ``A'' or ``B'' if one response is clearly better than the other.

Do not let the ordering or position of the responses influence your judgment; evaluate A and B solely on their content and quality.
}}
\caption{Example system prompt for the AI evaluator agent when the last student response has no textual response (facial expression reaction only).}
\label{fig:system_prompt_for_evaluator_agent_silent}
\end{figure*}

\begin{figure*}[t]
\centering
\fbox{\parbox{0.95\linewidth}{
You are an expert educational evaluator.

Your task is to compare two AI tutor responses to a student in a learning session and decide which one is better.

You must judge the responses along three independent dimensions:

\vspace{1em}

1. **pedagogical\_effectiveness** — Which response is clearer and more pedagogically effective?

\vspace{1em}

2. **empathetic\_responsiveness\_to\_facial\_expressions** — Which response shows greater awareness of, and responsiveness to, the student's emotional or cognitive state reflected in their facial expression?

\vspace{1em}

3. **empathetic\_responsiveness\_to\_textual\_responses** — Which response shows greater awareness of, and responsiveness to, the student's emotional or cognitive state reflected in their textual response?

\vspace{1em}

For each of the three evaluation questions, choose **Equal**, **A**, or **B**.  

Only choose ``A'' or ``B'' if one response is clearly better than the other.

Do not let the ordering or position of the responses influence your judgment; evaluate A and B solely on their content and quality.
}}
\caption{Example system prompt for the AI evaluator agent when the last student response includes a textual response.}
\label{fig:system_prompt_for_evaluator_agent_non_silent}
\end{figure*}

\begin{figure*}[t]
\centering
\fbox{\parbox{0.95\linewidth}{
A student is interacting with an AI tutor. Here is the conversation so far:

\textcolor{red}{\{conversation\_history\}}

Now, two candidate tutor responses are given below for the next turn.

Based on the conversation so far and the student’s facial expressions, which response is better?

\vspace{1em}

Response A:  

\textcolor{red}{\{response\_a\}}

\vspace{1em}

Response B:  

\textcolor{red}{\{response\_b\}}
}}
\caption{Example user input prompt for the AI evaluator agent}
\label{fig:input_prompt_for_evaluator_agent}
\end{figure*}

\end{document}